\documentclass[acmsmall,authorversion]{acmart}
\acmSubmissionID{1019}

\usepackage{tikz}
\usetikzlibrary{patterns}
\usetikzlibrary{arrows.meta}
\usetikzlibrary{fadings}
\usepackage{fancyvrb}
\usepackage{soul}
\usepackage{bm}
\usepackage{pgfplots}
\usepackage{pgfplotstable}
\usepgfplotslibrary{patchplots,colorbrewer,groupplots}
\usetikzlibrary{shadows,backgrounds,arrows,positioning,matrix,calc}
\pgfplotsset{compat=newest}
\usepackage[export]{adjustbox}
\usepackage{nicefrac}
\usepackage{gensymb}
\usepackage{graphicx}
\usepackage{gensymb}
\usepackage{color, colortbl}
\usepackage{stmaryrd} 
\usepackage{pifont}
\usepackage[figuresleft]{rotating}
\usepackage{enumitem}

\definecolor{myred}{rgb}{0.86,0.00,0.00}
\definecolor{myredlight}{rgb}{0.97,0.75,0.75}
\definecolor{myredlighter}{rgb}{0.99,0.94,0.94}
\definecolor{myredlighterr}{rgb}{1.0,0.98,0.98}
\definecolor{myblue}{rgb}{0.00,0.20,0.70}
\definecolor{mybluelight}{rgb}{0.75,0.80,0.93}
\definecolor{mybluelighter}{rgb}{0.94,0.95,0.98}
\definecolor{mybluelighterr}{rgb}{0.98,0.99,1.0}
\definecolor{mygreen}{rgb}{0.10,0.50,0.10}
\definecolor{mygreenlight}{rgb}{0.78,0.88,0.78}
\definecolor{mygreenlighter}{rgb}{0.94,0.97,0.94}
\definecolor{mygreenlighterr}{rgb}{0.99,0.99,0.99}
\definecolor{mygrey}{rgb}{0.40,0.40,0.40}
\definecolor{mygreylight}{rgb}{0.85,0.85,0.85}
\definecolor{mygreylighter}{rgb}{0.96,0.96,0.96}
\definecolor{mygreylighterr}{rgb}{0.99,0.99,0.99}
\definecolor{myorange}{rgb}{1.0,0.50,0.00}
\definecolor{myorangelight}{rgb}{1.0,0.87,0.75}
\definecolor{myorangelighter}{rgb}{1.0,0.96,0.93}
\definecolor{myorangelighterr}{rgb}{1.0,0.99,0.98}

\newcommand{\TwinID}{\textsc{Twin}}
\newcommand{\PrevID}{\textsc{Prev}}
\newcommand{\NextID}{\textsc{Next}}
\newcommand{\VertID}{\textsc{Vert}}
\newcommand{\EdgeID}{\textsc{Edge}}
\newcommand{\FaceID}{\textsc{Face}}

\usepackage{booktabs} 

\usepackage{algorithm}
\usepackage{algpseudocode}

\acmJournal{TOG}

\setcopyright{acmcopyright}
\acmJournal{PACMCGIT}
\acmYear{2022} \acmVolume{5} \acmNumber{3} \acmArticle{} \acmMonth{7} \acmPrice{15.00}\acmDOI{10.1145/3543868}

\acmDOI{doi.org/10.1145/3543868}

\begin{document}
\title{Htex: Per-Halfedge Texturing for Arbitrary Mesh Topologies}

\author{Wilhem Barbier}
\orcid{0000-0001-5208-5622}
\affiliation{%
 \institution{Unity Technologies}
 \country{France}
}
\email{wilhem@wbrbr.org}

\author{Jonathan Dupuy}
\orcid{0000-0002-4447-3147}
\affiliation{%
 \institution{Unity Technologies}
 \country{France}
}
\email{jonathan.dupuy@outlook.com}

\renewcommand\shortauthors{Barbier and Dupuy}

\begin{teaserfigure}
    \begin{center}
        \input{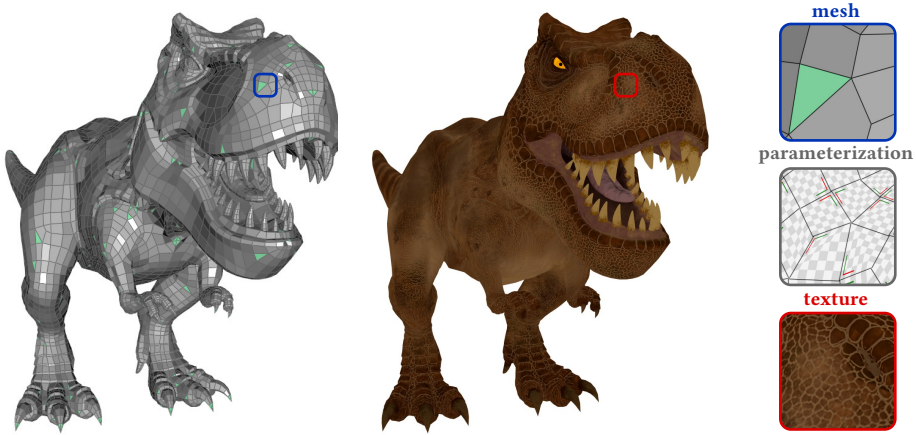}
    \end{center}
    \vspace{-0.0cm}
    \caption{\label{fig_teaser} 
    A production asset rendered with Htex textures. 
    Htex is a GPU-friendly alternative to Ptex and mesh colors that supports 
    non-quad topologies. This frees artists from UV-authoring during asset 
    creation, and yields high-quality seamless texture filtering. Due to its 
    GPU-friendly nature, Htex allows for fast rendering times through look development, 
    previzualization, and/or video-game prototyping. 
    }
\end{teaserfigure}

\begin{abstract}
We introduce per-halfedge texturing (Htex) a GPU-friendly method for 
texturing arbitrary polygon-meshes without an explicit parameterization. 
Htex builds upon the insight that halfedges encode an intrinsic 
triangulation for polygon meshes, where each halfedge spans a unique triangle 
with direct adjacency information. Rather than storing a separate texture 
per face of the input mesh as is done by previous parameterization-free 
texturing methods, 
Htex stores a square texture for each halfedge and its twin. 
We show that this simple change from face to halfedge induces two important 
properties for high performance parameterization-free texturing. 
First, Htex natively supports arbitrary polygons without requiring dedicated code 
for, e.g, non-quad faces. Second, Htex leads to a 
straightforward and efficient GPU implementation that uses only three 
texture-fetches per halfedge to produce continuous texturing across 
the entire mesh. 
We demonstrate the effectiveness of Htex by rendering 
production assets in real time. 
\end{abstract}

%


%
%


\maketitle

\clearpage

\section{Introduction}
\label{sec_intro}

\begin{figure*}[b]
    \begin{center}
        \input{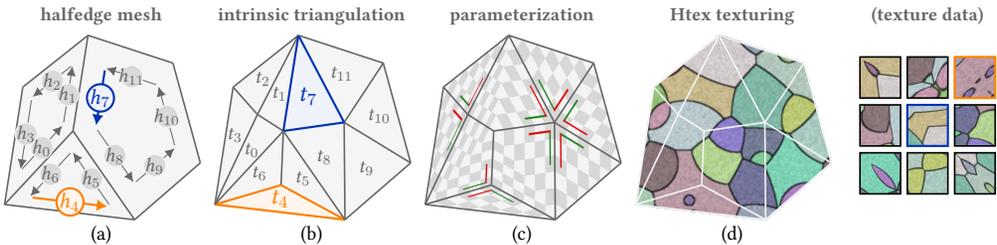} 
    \end{center}
    \vspace*{-0.0cm}
    \caption{\label{fig_htex} 
        Htex Overview. Htex builds upon (a) a halfedge mesh. 
        We make the key observation that halfedges and halfedge/twin pairs
        respectively define (b) an intrinsic triangulation, and (c) an intrinsic 
        quadrangulation of the mesh they belong to. The union of all halfedge/twin 
        pairs and boundary halfedges thus yields a new intrinsic mesh that covers the 
        entire surface of the original polygon mesh. Htex relies on (c) a simple 
        parameterization for each of the intrinsic face spanned by this union and 
        stores (d) a texture for each of them.
    }
\end{figure*}

\paragraph*{Context}
Per-face texturing (Ptex) is a ubiquitous texturing method for 3D polygon meshes:
it is used in offline production rendering as well as in content creation pipelines 
that depend on interactive modeling software~\cite{Burley2008}. 
Compared to conventional texturing, Ptex completely frees artists from 
UV-authoring while providing seamless filtering at the same time, thus enabling faster 
prototyping and artistic iteration. A key issue however is that Ptex can not be fully 
accelerated on the GPU\footnote{This is also true for its variant mesh 
colors~\cite{Yuksel2010} as we discuss in more detail in Section~\ref{sec_previous}.}. 
This can quickly hinder production pipelines because it becomes harder to guarantee 
interactive framerates as asset complexity grows through time for, e.g., modeling 
sessions or film pre-visualization. 

\paragraph*{Motivation for Arbitrary Topologies}
Ptex's GPU support is only partial in that existing implementations are restricted to 
quad-only meshes. While most assets tend to exhibit such topologies, many others 
will occasionally carry a few triangles and/or N-gons as shown in Figure~\ref{fig_teaser}.
In theory, such non-quad faces could be removed via automated and/or manual re-meshing. 
However, enforcing topological edits would greatly diminish the benefits that Ptex 
brings in terms of rapid iteration times (since re-meshing can be as tedious as UV-authoring). 
We therefore argue in favor of an alternative texturing method capable of retaining 
the aforementioned benefits of Ptex while adding seamless 
support for arbitrary polygons in a GPU implementation. We introduce such a method here, 
which we refer to as per-halfedge texturing (Htex).

\paragraph*{Contributions and Outline}
We arrived at Htex by first identifying that Ptex extensively relies on topological 
information. 
Based on this observation, we looked into using halfedges as a main texturing 
entity (as opposed to faces in the case of Ptex). 
We found that halfedges encode an intrinsic quadrangulation of their 
polygon-mesh, which can be used for a Ptex-like implementation. 
We build upon these novel results to devise Htex and describe how it leads to an 
efficient GPU implementation. The remainder of this article is summarized in 
Figure~\ref{fig_htex} and organized as follows:
\begin{itemize}
    \item In Section~\ref{sec_halfedges}, we present the intrinsic meshes we build solely 
    from halfedges.
    \item In Section~\ref{sec_htex}, we leverage these meshes to devise our novel 
    texturing method, which stores a square texture for each halfedge/twin pair.
    \item In Section~\ref{sec_results}, we provide rendering results and performance 
    measurements against conventional UV-mapping and a state-of-the-art Ptex GPU 
    implementation.
\end{itemize}

\section{Related Work}
\label{sec_previous}

In this section, we position our work with respect to the relevant literature.
Note that we do not discuss all texture mapping methods here as 
this is out of the scope of this paper. We refer the interested reader to the thorough survey of 
Yuksel~et~al.~\shortcite{Yuksel2019} for an overview of such methods. 

\paragraph*{UV Atlas}
The most ubiquitous method for texturing a polygon mesh consists in mapping 
it onto a square domain. In turn, 
this square domain is discretized into a texture called an atlas or simply 
UV~\cite{Maillot1993}, which can be edited by artists via any 2D
image-processing tool. The main issue with this approach is that the UV map
generally needs to satisfy an over-constrained set of problems. 
Such problems include (but are not restricted to) low distortion, 
symmetry optimizations, seam minimization, compliance with image-processing 
brushes and filters, etc. This makes it difficult to devise 
an automatic UV generation tool that always works. Consequently, UVs  
require manual tweaking by artists, which is a tedious and time-consuming 
process as mentioned in introduction. What is worse is that, in the general case, 
UV-mapping systematically produces seams that lead to texture-discontinuities on 
the mesh surface. 
Recently, Liu~et~al.~\shortcite{LiuFerguson2017} introduced an optimization that 
post-processes a given texture to make its data continuous at the mesh's surface. 
However, we have found that this optimization leaves a few discontinuities.
While this is fine for most textures, it makes it impossible to use UVs for 
displacement maps, as shown in Figure~\ref{fig_dmap}.
Htex is free from such limitations and allows to do displacement mapping seamlessly.

%

%

\paragraph*{Ptex}
Ptex~\cite{Burley2008} addresses the issue of quad-mesh parameterization for texturing 
by assigning a texture per quad. In Ptex, texture filtering is achieved per-quad by 
systematically sampling its associated texture as well as those opposite to its edges. 
This approach produces seamless filtering within each quad as well as at its boundaries.

\paragraph*{Ptex on the GPU}
Several GPU implementations of Ptex have been proposed. Early implementations 
relied on GPUs that were limited in the number of simultaneous texture-reads 
they could perform. To alleviate this issue, Ptex textures can be tightly packed 
into a larger texture~\cite{McDonald2011,Kim2011,Niessner2013Displacement}. Unfortunately, 
this makes filtering challenging as filter taps can erroneously sample 
neighboring data, especially when anisotropic filtering is enabled.
Borders can be used to mitigate the issue, but this never completely solves the 
problem and can add a significant memory overhead. More recent implementations 
no longer attempt to minimize the number of texture samplers and rely on 
texture arrays of bindless textures to sample directly from the Ptex 
textures. For filtering this data, McDonald introduces borderless 
Ptex~\cite{McDonald2013}. Htex heavily builds upon this 
implementation so we discuss it more thoroughly in Section~\ref{sec_htex_implementation}.
For the sake of completeness, we also mention the implementation of 
Toth~\shortcite{Toth2013}, which avoids sampling neighboring quads by 
discarding filter taps and relying on MSAA to reconstruct a smooth signal.
This approach produces good filtering quality with high MSAA factors, but does 
not allow texture magnification as opposed to borderless Ptex. 

\paragraph*{Mesh Colors}
In the case of quad-only meshes, mesh colors~\cite{Yuksel2010} provides a variant of 
Ptex that uses the dual parameterization of Ptex. The advantage of this approach is 
that it alleviates the need to sample from the neighbors of each quad, because 
boundary data is duplicated by construction. However, it renders its GPU implementation 
more difficult as the required logic for performing trilinear and anisotropic 
filtering no longer maps to GPU texturing units~\cite{Yuksel2017}. In order to 
overcome these issues, some well-identified modifications have to be made to 
existing hardware~\cite{Mallett2019,Mallett2020}. In this work, we target maximum 
compatibility with current hardware and therefore avoid the dual parameterization 
of mesh colors entirely. Nevertheless, nothing prevents Htex to be used with 
the dual parameterization of mesh colors.
 
\paragraph*{Issues with Non-Quads}
A common limitation of existing GPU implementations of Ptex is that they do not support 
non-quad faces. This is because the way non-quad faces are handled is problematic 
for a GPU: Ptex splits them into sub-quads that require dedicated filtering 
code~\cite{Disney2022}. While this approach works fine on CPU-based implementations, 
it constitutes a challenge for a GPU-based implementation due to the algorithmic 
branching it fundamentally incurs. We argue that the main reason for this divergent 
behavior is due to the fact that Ptex was originally built for quad-only meshes.
Before elaborating Htex, we considered non-quad support as mandatory. As a result, 
Htex natively supports non-quads on the GPU. The key that makes Htex work is that 
it does not store a texture per face of the polygon mesh, but on an intrinsic 
quadrangulation. As we will show in the next section, this quadrangulation
arises naturally when using a halfedge mesh and does not need to be stored 
explicitly in memory.

%
%

\paragraph*{Volumetric Texturing}
Volumetric textures are an alternative method to per-face texture mapping methods,
which use the 3D positions of the mesh surface as texture 
coordinates~\cite{Benson2002,DeBry2002,Dolonius2020}.
The key advantages of Htex over volumetric texturing are its efficiency and ease of 
implementation: As we will show, Htex maps to current GPU texture filtering hardware 
natively and does not require any other data than a halfedge mesh, which we present in the next 
paragraph. In contrast, volumetric textures require 3D (rather than 2D) logic for 
sampling along with a sparse data-structure for storage optimizations, both of which 
must be implemented in software. Another important limitation of volumetric textures over Htex is 
their current lack of support for anisotropic filtering.

\begin{figure*}
        \begin{tikzpicture}
    \begin{scope}[scale = 1.0]

    \begin{scope}[shift={(0, 0)}]
        \begin{scope}[scale=3.0]
            \input{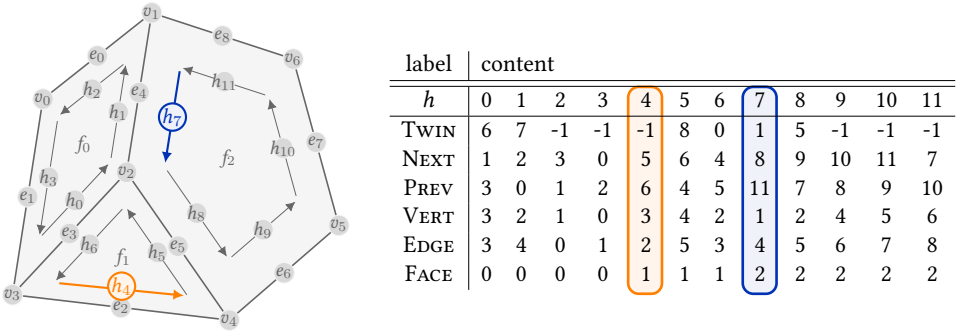}
        \end{scope}

        \begin{scope}[shift={(6.6, 0.5)}]
            \fill[mybluelighter, rounded corners] (0.975, -1.6) rectangle ++(0.45, 2.72);
            \fill[myorangelighter, rounded corners] (-0.55, -1.62) rectangle ++(0.45, 2.72);
            \draw[myblue, line width = 1.0, rounded corners] (0.975, -1.62) rectangle ++(0.45, 2.72);
            \draw[myorange, line width = 1.0, rounded corners] (-0.55, -1.62) rectangle ++(0.45, 2.72);
            \node (tbl) {
                {\scalebox{0.9}{
                \begin{tabular}{c|cccccccccccc}
                    label & \multicolumn{12}{l}{content} \\
                    \hline
                    \hline
                    $h$ & 0 & 1 & 2 & 3 & 4 & 5 & 6 & 7 & 8 & 9 & 10 & 11 \\
                    \hline
                    \TwinID & 6 & 7 & -1 & -1 & -1 & 8 & 0 & 1 & 5 & -1 & -1 & -1 \\
                \NextID & 1 & 2 & 3 & 0 & 5 & 6 & 4 & 8 & 9 & 10 & 11 & 7 \\
                \PrevID & 3 & 0 & 1 & 2 & 6 & 4 & 5 & 11 & 7 & 8 & 9 & 10\\
                \VertID & 3 & 2 & 1 & 0 & 3 & 4 & 2 & 1 & 2 & 4 & 5 & 6 \\
                \EdgeID & 3 & 4 & 0 & 1 & 2 & 5 & 3 & 4 & 5 & 6 & 7 & 8 \\
                \FaceID & 0 & 0 & 0 & 0 & 1 & 1 & 1 & 2 & 2 & 2 & 2 & 2 \\
            \end{tabular}}}
            };
            
        \end{scope}
    \end{scope}

\end{scope}
\end{tikzpicture} 
    \caption{\label{fig_halfedge}
        Halfedge mesh representation used for the implementation of Htex. The blue and orange 
        halfedges are highlighted as illustrative examples of a regular and border halfedge, respectively. 
    }
\end{figure*}

\paragraph*{Halfedge Mesh Representation}
Htex builds upon a halfedge mesh. A halfedge 
mesh~\cite{Halfedge1985,Halfedge1999,PolygonMeshProcessing2010} decomposes a polygon mesh 
into two sets: a set of vertex points and a set of halfedges. The 
vertex points encode the positional information for the mesh, while the 
halfedges--from which edges and faces emanate naturally--encode its topology.
The halfedge are solely characterized by their operators \TwinID, \NextID, \PrevID, 
\VertID, \EdgeID, and \FaceID. In our implementation, we rely on a generalization of 
directed edges~\cite{Directed1998} as our halfedge data-structure, which is 
illustrated in Figure~\ref{fig_halfedge} along with the aforementioned halfedge operators. 
Note that this same data-structure was used 
recently in the context of GPU-based subdivision~\cite{Dupuy2021}.

\section{Intrinsic Meshing with Halfedges}
\label{sec_halfedges}

\begin{figure*}
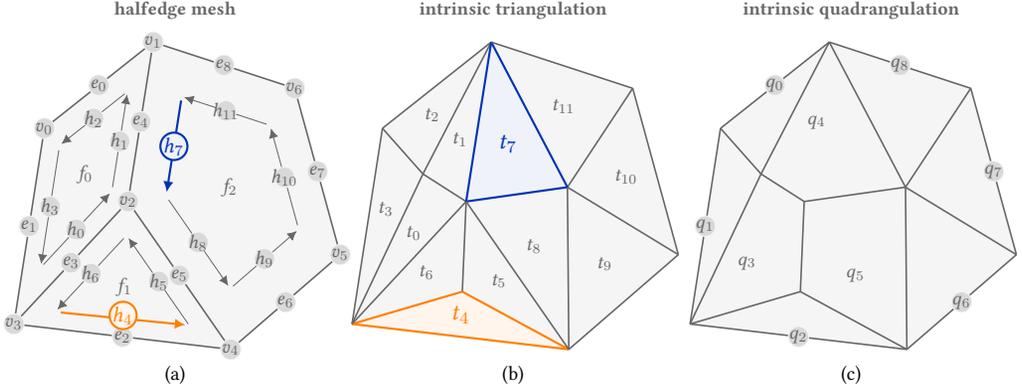

    \begin{center}
        \begin{tikzpicture}
    \begin{scope}[shift={(0.0, 0.0)}]
        \begin{scope}[shift={(+0.0, -0.35)}, scale=3.0]
            \node[mygrey, draw=none, anchor=center] 
            at (0.0, 1.0) {\scalebox{0.7}{\textbf{halfedge mesh}}};
                \input{figures/figure-halfedge-mesh.tex}
            \node[black, draw=none, anchor=center] at (0.0, -0.6) {\scalebox{0.75}{(a)}};
        \end{scope}

        \begin{scope}[shift={(4.5, -0.35)}, scale=3.0]
            \node[mygrey, draw=none, anchor=center] 
            at (0.0, 1.0) {\scalebox{0.7}{\textbf{intrinsic triangulation}}};
            \input{figures/figure-htex-triangulation.tex}
            \node[black, draw=none, anchor=center] at (0.0, -0.6) {\scalebox{0.75}{(b)}};
        \end{scope}
        
        \begin{scope}[shift={(9.0, -0.35)}, scale=3.0]
            \node[mygrey, draw=none, anchor=center] 
            at (0.0, 1.0) {\scalebox{0.7}{\textbf{intrinsic quadrangulation}}};
            \input{figures/figure-htex-parameterization-detail.tex}
            \node[black, draw=none, anchor=center] at (0.0, -0.6) {\scalebox{0.75}{(c)}};
        \end{scope}

    \end{scope}
\end{tikzpicture} 
    \end{center}
    \vspace*{-0.0cm}
    \caption{\label{fig_intrinsic} 
    Intrinsic triangulation and quadrangulation with halfedges.
    }
\end{figure*}

In this section, we describe the key insight that Htex builds upon.
Specifically, we show that halfedges encode an intrinsic triangulation of 
the polygon mesh they belong to (Section~\ref{sec_halfedges_triangles}). 
We then show how to further pair such triangles to form an intrinsic quadrangulation
over the polygon mesh (Section~\ref{sec_halfedges_quads}).

\subsection{Halfedges as an Intrinsic Triangulation}
\label{sec_halfedges_triangles}

The intrinsic triangulation we introduce is shown in 
Figure~\ref{fig_intrinsic}~(a,b). Its properties are straightforward:

\paragraph*{A Triangle for Each Halfedge}
Each halfedge maps to exactly one triangle. For instance:
\begin{itemize}
    \item[-] the blue halfedge $h_7$ 
    in Figure~\ref{fig_intrinsic}~(a) maps to the blue triangle $t_7$ in 
    Figure~\ref{fig_intrinsic}~(b). 
    \item[-]  the orange halfedge $h_4$ 
    in Figure~\ref{fig_intrinsic}~(a), maps to the orange triangle $t_4$ in 
    Figure~\ref{fig_intrinsic}~(b).
\end{itemize}

\paragraph*{Triangle Vertices}
We retrieve the vertices of each triangle thanks to the halfedge 
operators \NextID~and \PrevID. For instance, the vertices of the blue triangle 
$t_7$ in Figure~\ref{fig_intrinsic}~(b) are 
\begin{align*}
    \text{the vertex } v_0 &:= \VertID(\NextID(h_7)) = \VertID(h_8),\\
    \text{the average } 
    v_1 &:= \frac{1}{5} \left(\VertID(h_7) + \VertID(h_8) + \cdots + \VertID(h_{11}) \right), \\
    \text{and the vertex }  v_2 &:= \VertID(h_7).
\end{align*}
Algorithm~\ref{alg_HtexVertices} provides pseudocode for computing these vertices.

\paragraph*{Triangle Adjacency}
We retrieve adjacent triangles thanks to the 
halfedge operators \NextID, \PrevID, and \TwinID. For instance:
\begin{itemize}
    \item[-] the triangles adjacent to that of the blue triangle $t_7$ in Figure~\ref{fig_intrinsic}~(b) are 
    those formed by halfedges \mbox{$\PrevID(h_7) = h_{11}$}, \mbox{$\NextID(h_7) = h_{8}$}, 
    and \mbox{$\TwinID(h_7) = h_1$}.
    \item[-] the triangles adjacent to that of the orange triangle $t_4$ in 
    Figure~\ref{fig_intrinsic}~(b) are those formed by halfedges \mbox{$\PrevID(h_4) = h_{6}$}, 
    \mbox{$\NextID(h_4) = h_{5}$}, and \mbox{$\TwinID(h_4) = -1$}, since $h_4$ is a 
    boundary halfedge.
\end{itemize} 
Note that by construction, all triangles have exactly three neighbors except for those 
lying at the mesh boundary, which have exactly two. In the latter case, the neighbors 
are always given by the boundary halfedge's $\PrevID$ and $\NextID$ operators.

\begin{algorithm}
    \small  
    \begin{algorithmic}[1]
        \caption{Halfedge to triangle}
        \label{alg_HtexVertices}
        \Function{IntrinsicTriangleVertices}{halfedgeID: integer}
        \State nextID $\gets$ \Call{Next}{halfedgeID}
        \State $v_0$ $\gets$ \Call{Vert}{nextID}
        \State $v_2$ $\gets$ \Call{Vert}{halfedgeID}
        \State $v_1$ $\gets$ $v_2$
        \State $n$ $\gets$ 1
        \State $h$ $\gets$ nextID
        \Statex
        \While{$h$ $\neq$ halfedgeID}
        \State $v_1$ $\gets$ $v_1$ $+$ \Call{Vert}{$h$}
        \State $n$ $\gets$ $n + 1$
        \State $h$ $\gets$ \Call{Next}{$h$}
        \EndWhile
        \Statex
        \State $v_1$ $\gets$ $v_1 / n$
        \Statex
        \State \Return $\langle v_0,\, v_1,\,v_2\rangle$
        \EndFunction
    \end{algorithmic}
\end{algorithm}

\subsection{Building an Intrinsic Quadrangulation with Triangle Pairs}
\label{sec_halfedges_quads}

We now describe an intrinsic quadrangulation that builds upon the triangulation described 
in the previous subsection. We start by considering the case of boundary-free polygon-meshes, 
and then generalize our result for boundaries.


\paragraph*{Quads as Halfedge/Twin Pairs}
In the case of a boundary-free polygon mesh, we have $E = 2H$, where $E$ and $H$ respectively 
denote its number of edges and halfedges. This suggests that 
we can build a quadrangulation of $E$ quads by creating unique pairs of adjacent 
triangles. In practice, we form an intrinsic quad by pairing the two triangles formed 
by a halfedge and its twin. The advantage of this approach is that it 
creates a bijection between the edges of the polygon mesh and its intrinsic 
quadrangulation. A practical consequence of this bijection is that each halfedge 
then maps to the intrinsic quad it spans using its $\EdgeID$ operator.
Our approach is illustrated in Figure~\ref{fig_intrinsic}, where we pair the 
blue triangle $t_7$ with its twin $t_1$ in (\ref{fig_intrinsic},~a) to 
yield the quad $q_4$ in (\ref{fig_intrinsic},~c). From halfedge perspective, 
the halfedge pair $h_1$ and $h_7$ form the edge 
\mbox{$\EdgeID(h_7) = \EdgeID(h_1) = e_4$}, which maps to quad $q_4$. 
Notice that we purposely use consistent subscripts between the edges $e$ 
in (\ref{fig_intrinsic},~a) and the quads $q$ in (\ref{fig_intrinsic},~c) to emphasize 
their bijective relation.



\paragraph*{Boundaries}
We now generalize our intrinsic quadrangulation in the case where boundaries are present
in the polygon-mesh. In such configurations, we still form a quad for 
each non-boundary edge. This leaves isolated triangles at the boundary as shown in 
Figure~\ref{fig_intrinsic}. We convert these isolated triangles into quads by 
pairing the boundary halfedges that span them with themselves. 
Intuitively, this is equivalent to placing a virtual halfedge on the other side of 
the boundary. We illustrate this approach using Figure~\ref{fig_intrinsic}: we build 
the quad $q_2$ in (\ref{fig_intrinsic},~c) by pairing the orange halfedge $h_4$ in 
(\ref{fig_intrinsic},~a) with itself. 
Again, we emphasize the bijection between edges $e$ in (\ref{fig_intrinsic},~a) and 
quads $q$ in (\ref{fig_intrinsic},~c) by using consistent subscripts. 
Our construction thus forms $E \in [H,~2H)$ intrinsic quads for any non-manifold 
mesh. Furthermore, each halfedge still maps to the intrinsic quad it spans via its 
$\EdgeID$ operator. For instance, the orange boundary halfedge $h_4$ in 
(\ref{fig_intrinsic},~a) maps to the quad $q_2$ in (\ref{fig_intrinsic},~c) via its 
edge $e_2 = \EdgeID(h_4)$.

\section{Per-Halfedge Texturing}
\label{sec_htex}

In this section, we leverage the intrinsic mesh we derived in 
Section~\ref{sec_halfedges} to devise Htex. We start by a high-level description 
of Htex (Section~\ref{sec_htex_overview}). Next, we build our mesh 
parameterization (Section~\ref{sec_htex_parameterization}). We then further describe
how we store and filter texture data (Section~\ref{sec_htex_storage}). Finally, we 
describe our GPU implementation, which is based on 
OpenGL (Section~\ref{sec_htex_implementation}).


\subsection{High-Level Description}
\label{sec_htex_overview}
Htex stores a separate texture per intrinsic quad of the halfedge mesh, each of which 
can be independently sized. 
Since we are now dealing with quad-only topologies, we can use the same approach as 
Ptex for texture filtering. However, Htex offers a few additional advantages.
For filtering, we look-up adjacent quads similarly to 
Ptex. Since all adjacency information is provided directly by the halfedges of 
the mesh, Htex only stores texel data without any further adjacent information. 
For rendering, we draw the intrinsic 
triangulation, i.e., a triangle for each halfedge rather than the actual polygons of 
the mesh. Note that our triangulation does not alter the shape of the original 
polygonal mesh in any way due to its construction. We then perform texture lookup using 
halfedge operators, which tell us which textures to lookup in memory for seamless filtering. 



\subsection{Parameterization}
\label{sec_htex_parameterization}

Htex stores a texture for each of our intrinsic quad and we derive here the 
parameterization that tells how to map texels onto the actual surface of the 
polygon mesh. To this end, we use the fact that, by construction, each quad can be 
triangulated into the two intrinsic triangles formed by a halfedge and its twin 
(or itself for a boundary halfedge). We therefore parameterize each quad as the 
union of these two intrinsic triangles as follows:

\paragraph*{Intrinsic Triangle Parameterization}
We start with a parameterization for our intrinsic triangles using barycentric-coordinate 
space \mbox{$(u, v, w)$}. We configure this space such that:
\begin{enumerate}[label=(\roman*)]
    \item $(u, v, w) = (1, 0, 0)$ maps to the vertex point $v_2 = \VertID(h)$, and
    \item $(u, v, w) = (0, 1, 0)$ maps to the vertex point $v_0 = \VertID(\NextID(h))$, 
\end{enumerate}
where $h$ denotes the halfedge that spans the intrinsic triangle.

\paragraph*{Retrieving the Quad Texture Coordinates}
We define our quad parameterization as a split into that of the two intrinsic 
triangles that compose it along the diagonal (1,0) -- (0,1) in uv-space. 
We determine whether to map 
a triangle on the upper or lower part of this split using the fact that we represent 
halfedges as integers: We map the halfedge with largest integer value to 
the lower left part of the quad. Our approach is illustrated in 
Figure~\ref{fig_htex}~(c), where each frame is located at the origin of the quad's 
parametric domain. Thanks to this construction, we can map the parameterization of 
an intrinsic triangle into that of a quad. Algorithm~\ref{alg_QuadParameterization} 
provides pseudocode for computing such a mapping given a halfedge and the barycentric 
coordinates of its associated intrinsic triangle. 

\begin{algorithm}
    \small
    \begin{algorithmic}[1]
        \caption{Intrinsic Triangle to Texture Coordinates}
        \label{alg_QuadParameterization}
        \Function{TriangleToQuadUV}{
            halfedgeID: integer, $\langle u, v\rangle$: barycentric coordinates
        }
        \If{halfedgeID $>$ \Call{Twin}{halfedgeID}}
            \State \Return $\langle u, v\rangle$
        \Else
            \State \Return $\langle 1-u, 1-v\rangle$
        \EndIf
        \EndFunction
    \end{algorithmic}
\end{algorithm}

\paragraph*{Rendering the Intrinsic Triangulation}
Our parameterization allows to render a polygon-mesh using its intrinsic triangulation.
In order to texture these triangles, we only need the halfedge that spans it as 
it provides all the operators to retrieve its associated texel data.

\subsection{Storage and Filtering}
\label{sec_htex_storage}

\paragraph*{Texture Data}
We store texture data as regular textures, just like Ptex. In the case of 
boundary halfedges, our construction only fills half of the texture, 
effectively resulting in wasted space at mesh boundaries. Fortunately, 
boundaries are very sparse within typical polygon meshes. As such, 
we did not make any effort to avoid this waste. Rather, we mirror the 
texel content of the existing halfedge as shown in 
Figure~\ref{fig_htex}~(d) (see the orange inset), which is useful for 
the seamless filtering approach we introduce next. In terms of 
texture count, Htex stores exactly $E$ textures, where $E$ denotes 
the number of edges on a mesh. Compared to Ptex in the case of a quad-only mesh, 
Htex thus stores a factor of $E/F$ more textures, where $F$ denotes the number 
of quads in the mesh. For non-quad faces, Htex stores as many textures as Ptex.

\paragraph*{Filtering}
Our filtering approach works with standard GPU texture filtering units.
In essence, it is similar to that of Ptex as it requires fetching the texel 
content of neighboring faces. An important distinction however is that, since we render 
the intrinsic triangulation rather than the actual polygon mesh, we only need to deal 
with 3 neighbors (rather than 4 for Ptex). 
As explained in Section~\ref{sec_halfedges}, we retrieve these neighbors using the 
halfedge operators $\NextID$, $\PrevID$, and $\TwinID$. Furthermore, since we store 
halfedges and their twins together in the same texture, sampling from 
the intrinsic quad already takes care of one of the neighbors. This leaves two 
extra lookups for $\PrevID$ and $\NextID$. Htex thus provides the same filtering 
capabilities as Ptex but at the cost of 3 texture fetches rather than 5.

\subsection{GPU Implementation}
\label{sec_htex_implementation}

Our GPU implementation of Htex is very similar to borderless Ptex~\cite{McDonald2013}.
We provide more details in the following paragraphs.

\paragraph*{Texture and Sampler Initialization}
We store each texture in a separate texture object that we manipulate using the 
bindless texture extension \texttt{GL\_ARB\_bindless\_texture}. We sample these 
textures in the same fashion as borderless Ptex, which requires two main steps:
\begin{enumerate}
    \item set the sampler of each texture to \texttt{CLAMP\_TO\_BORDER} with a border 
    color of zero.
    \item add an extra border-normalization channel per texture resolution with all texels 
    set to one.
\end{enumerate}

\paragraph*{Rendering}
During rendering, we render each intrinsic triangle associated with a halfedge of 
the polygon mesh. For texture sampling, we proceed as shown in 
Algorithm~\ref{alg_Htexture}: Each triangle accumulates the sample of its texture with 
those associated with the halfedges $\NextID$ and $\PrevID$. For each of these three 
texture fetches, we also fetch the border-normalization channel and use it as 
normalization. An important step of Algorithm~\ref{alg_Htexture} is to convert 
the barycentric coordinates of the current triangles to its two neighbors.
Borderless Ptex accomplishes this by precomputing 16 transformation matrices. 
In our case, our intrinsic construction makes the transformation obvious and 
simply consist in reflecting the coordinates (see lines 7 and 9 in 
Algorithm~\ref{alg_Htexture}).

\paragraph*{Corner Pre-processing}
As borderless Ptex, Htex produces continuous filtering everywhere except at corner 
texels. In practice, this discontinuity is un-noticeable for most texture data. 
A notable exception is displacement mapping as the discontinuity results in cracks. 
For this case we follow previous methods~\cite{McDonald2011,Niessner2013Displacement}
and pre-process texel corners to enforce that they all have the same value.
Note that this produces continuous filtering whenever texture have the same
resolution. In the case of varying texture resolutions, we have observed that
a few cracks may still appear due to precision issues within GPU texture samplers.

\setlength{\textfloatsep}{12pt}
\begin{algorithm}
    \small
    \begin{algorithmic}[1]
        \caption{Htex Sampling on the GPU}
        \label{alg_Htexture}
        \Function{Htexture}{
            halfedgeID: integer, $\langle u, v\rangle$: barycentric coordinates
        }
        \State nextID $\gets$ \Call{Next}{halfedgeID}
        \State prevID $\gets$ \Call{Prev}{halfedgeID}

        \State $c = \langle c_0, \cdots, c_n\rangle$ $\gets$ $0$ \Comment{we have $n$ channels plus the border-normalization channel}

        \State $\langle x, y\rangle$ $\gets$ \Call{TriangleToQuadUV}{halfedgeID, $\langle u, v\rangle$}
        \State $c$ $\gets$ $c$ $+$ \Call{Texture}{\EdgeID(halfedgeID), $\langle x, y\rangle$}
        
        \State $\langle x, y\rangle$ $\gets$ \Call{TriangleToQuadUV}{nextID, $\langle v, -u\rangle$}
        \State $c$ $\gets$ $c$ $+$ \Call{Texture}{\EdgeID(nextID), $\langle x, y\rangle$}

        \State $\langle x, y\rangle$ $\gets$ \Call{TriangleToQuadUV}{prevID, $\langle -v, u\rangle$}
        \State $c$ $\gets$ $c$ $+$ \Call{Texture}{\EdgeID(prevID), $\langle x, y\rangle$}
        
        \State \Return $\langle \frac{c_0}{c_n}, \cdots, \frac{c_{n-1}}{c_n}\rangle$
        \EndFunction
    \end{algorithmic}
\end{algorithm}
\section{Results and Evaluation}
\label{sec_results}

In this section, we evaluate the performances of Htex. We first showcase some 
renderings to evaluate the filtering quality and scalability of Htex 
(Section~\ref{sec_results_rendering}). We then provide some performance 
measurements and compare them against borderless Ptex and classic 
UV-mapping (Section~\ref{sec_results_performances}). Finally, we discuss 
some limitations of Htex (Section~\ref{sec_results_limitations}).

\subsection{Rendering Results}
\label{sec_results_rendering}

\begin{figure*}
        \input{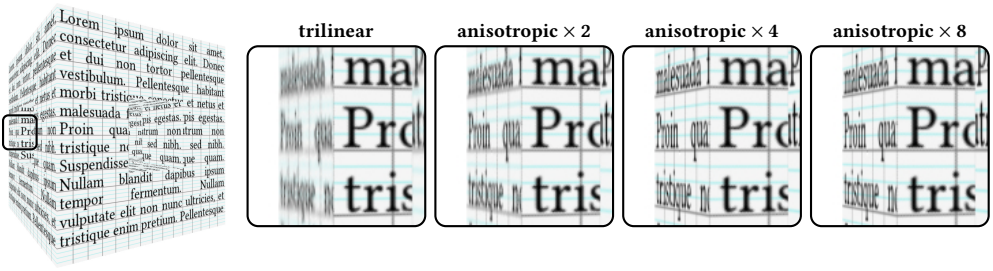} 
    \caption{\label{fig_filtering}
        Hardware anisotropic filtering with Htex.
    }
\end{figure*}

\paragraph*{Anisotropic Filtering}
Htex maps seamlessly to the GPU's hardware accelerated anisotropic filter. 
We demonstrate this in Figure~\ref{fig_filtering}, where we render an Htex asset with 
varying anisotropic filtering factors. We generated this asset by taking direct 
inspiration from the original Ptex article, see \cite{Burley2008},~Figure~(10).
To do so, we applied the same texture with tri-planar mapping on the original mesh, 
which is available on the Ptex website. Notice how anisotropic filtering improves the 
quality of the texture at grazing angles.
To further demonstrate Htex's filtering quality, we provide an animated rendering of the 
same asset in our supplemental video.

\begin{figure*}
        \input{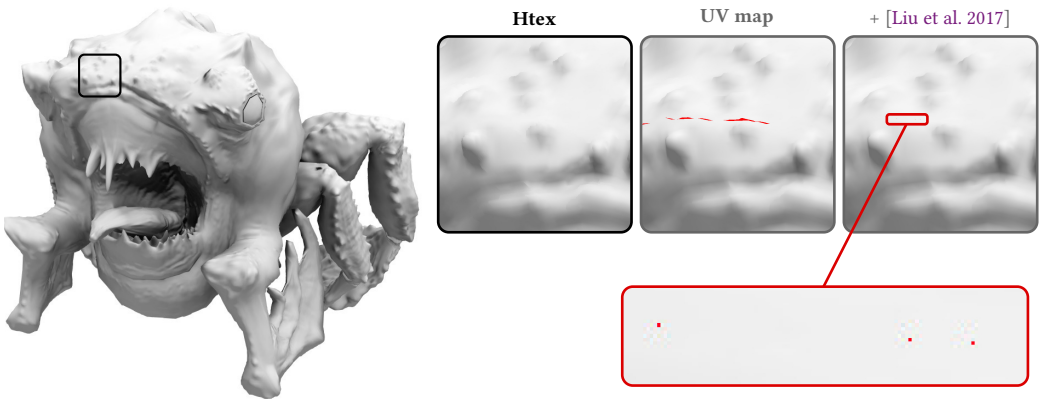} 
    \caption{\label{fig_dmap}
        Displacement mapping with Htex. Htex allows for artifact-free displacement 
        mapping on the GPU while UV maps will systematically produce cracks 
        (highlighted in red in insets) at UV-seams. Cracks can be greatly 
        reduced by the method of Liu~et~al.~\shortcite{LiuFerguson2017} although 
        never entirely (bottom inset).
    }
\end{figure*}

\paragraph*{Displacement Mapping}
Htex allows for crack-free displacement maps as illustrated in Figure~\ref{fig_dmap}. 
We emphasize that crack-free displacement is 
impossible to achieve with UV-maps as seams will systematically result in texture 
discontinuities. 
Furthermore, we mention that the asset used for 
this particular example is similar in nature to the one shown in Figure~\ref{fig_teaser}: 
it is composed of mostly quads (1494 in total), but also a few triangles (38 in total). 
Htex supports it effortlessly. Note that we further demonstrate the benefits of 
Htex with respect to UV-maps in the context of GPU-accelerated displacement mapping
in our supplemental video. 

\paragraph*{Scalability}
To demonstrate the scalability of Htex, we render production assets
in Figure~\ref{fig_teaser} and Figure~\ref{fig_ptex}. The asset from 
Figure~\ref{fig_teaser} consists of 10634 quads, 436 triangles, and 
562 boundary edges. The asset from Figure~\ref{fig_ptex} is a quad-only Ptex asset
with 5812 faces and 572 boundary edges. To produce the rendering in Figure~\ref{fig_ptex}, 
we resampled the Ptex textures into Htex ones, which consist of displacement and 
albedo data. We provide an animated zoom-in sequence in our supplemental video that 
relies on adaptive tessellation to render the asset in real time.
Note that we displace the mesh along the normals of the input mesh, 
rather than from those of the limit surface of Catmull-Clark subdivision. 
We plan to add subdivision in our implementation, but note that this is orthogonal 
to the problem of texturing, which Htex addresses. 

\subsection{Performances}
\label{sec_results_performances}

\paragraph*{Methodology}
We position Htex's performances against our implementation of borderless 
Ptex~\cite{McDonald2013} and conventional UVs. Our performance measurements 
correspond to the median rendering timing over 100 successive frames rendered 
at $1920\times1080$ resolution.
Our GPU rendering pipeline uses tessellation shaders and we measure 
its performances with and without sampling a displacement map. The fragment 
shader simply samples an extra RGB texture. The rendering involves a single asset, 
which we draw so that it occupies the entire framebuffer. 
Our test machine has an NVIDIA~RTX~3090~GPU and an Intel~i9-10980XE~CPU with 
128GiB of RAM.

\begin{table}[h]
    \begin{center}

 {
    \begin{tabular}{c|c|c} 
    & RGB only & RGB + disp \\ 
    \hline 
    UVs      & 0.40 ms  & 0.59 ms \\
    Htex     & 0.42 ms  & 0.68 ms \\
    speedup & $\times 0.95$  & $\times 0.87$ 
    \end{tabular}
}
    \end{center}
    \caption{Htex performance comparison against UVs on an NVIDIA RTX 3090.}
    \label{tab_perf_uv}
\end{table}

\paragraph*{Comparison Against UV-maps}
We report our timings for Htex's performance against UVs in Table~\ref{tab_perf_uv}. 
We performed our measurement using the asset from Figure~\ref{fig_teaser} for which 
we also have a UV map. As demonstrated by the reported numbers, Htex's overhead 
is almost negligible when displacement is disabled. As for when displacement is 
enabled, Htex's overhead becomes slightly more important.
We attribute this increased difference to the fact that texture fetches 
during vertex processing do not necessarily pipeline as well as in fragment 
shaders and cache misses are more frequent. We finally mention that 
the displaced, UV-based configuration is purely synthetic as UVs will systematically produce 
cracks on the final surface. As such, these numbers should rather be seen as 
a baseline.

\begin{table}[h]
    \begin{center}

 {
    \begin{tabular}{c|c|c} 
    & RGB only & RGB + disp \\ 
    \hline 
    Ptex      & 0.9 ms  & 2.4 ms \\
    Htex & 0.77 ms  & 1.7 ms \\
    speedup & $\times 1.17$  & $\times 1.41$ 
    \end{tabular}
}
    \end{center}
    \caption{Htex performance comparison against borderless Ptex~\cite{McDonald2013} on 
    an NVIDIA RTX 3090.}
    \label{tab_perf_ptex}
\end{table}

\begin{figure*}
        \input{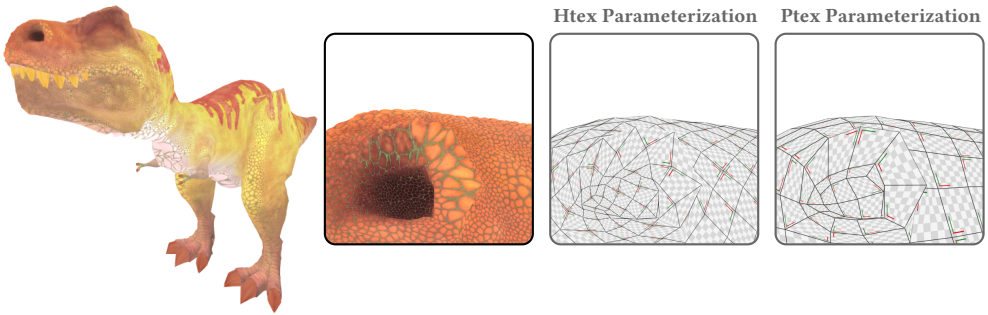} 
    \caption{\label{fig_ptex}
        Ptex production asset (Ptex T-rex model \copyright Walt Disney Animation Studios.) 
        rendered with Htex and parameterization comparison. 
    }
\end{figure*}

\paragraph*{Comparison Against Borderless Ptex}
We report our timings for Htex's performance against Ptex in Table~\ref{tab_perf_ptex}. 
We performed our measurement using the asset from Figure~\ref{fig_ptex}. As demonstrated 
by the reported numbers, Htex provides a solid speedup over our borderless Ptex implementation.
This is due to the fact that Htex requires less texture taps and memory accesses as 
we do not require fetching adjacency information or UV transformation matrices. We 
also re-emphasize that our implementation supports arbitrary polygons rather than just 
quads in the case of borderless Ptex.

\paragraph*{Memory Consumption}
The memory consumption of Htex is a function of the number of texels stored on the surface 
of the polygon mesh. Just like borderless Ptex, we also require an extra 
border-normalization channel per texture resolution for filtering. We mention that, 
in practice, the overhead incurred by this extra channel quickly becomes negligible as 
the number of channels increases. Moreover, we believe that the need for an extra 
border-normalization channel could be removed entirely by having GPU texture samplers 
return the blending value of each border. It remains to be discussed whether this is a 
reasonable possibility.

\subsection{Limitations}
\label{sec_results_limitations}


\begin{figure*}
        \input{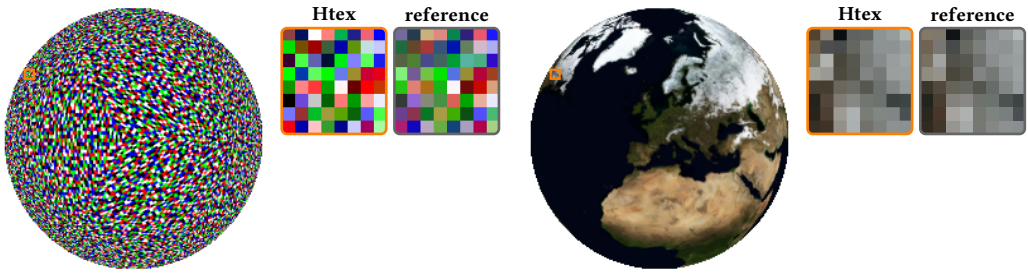} 
    \caption{\label{fig_aliasing}
        Htex texture aliasing. Htex aliases at the same resolution as its intrinsic quadrangulation.
    }
\end{figure*}

\paragraph*{Subpixel-Face-Filtering}
Fundamentally, Htex remains a per-face texturing solution. As such, it shares the same 
limitations as other per-face methods like Ptex and mesh colors. One such 
limitation is the fact that both texture and geometry alias at the same resolution.
This means that Htex is unable to filter texture data below the resolution of the 
intrinsic quadrangulation of its input mesh. We illustrate this issue in 
Figure~\ref{fig_aliasing}, where we render a highly tessellated sphere with Htex. 
Notice the aliasing at the silhouette of the sphere, which is especially visible for 
high-frequency textures. 


\paragraph*{Low-Quality Meshes}
Another limitation common to per-face texturing methods is that parameterization quality directly 
depends on that of the input polygon-mesh. If the input polygon-mesh carries, e.g., highly elongated faces, 
then so will the intrinsic quadrangulation and hence our parameterization will produce highly 
anisotropic texels. Note that per-face texturing methods can somewhat minimize this effect through 
their support for arbitrary-sized textures, although this will hardly be as good as having a high-quality 
mesh from the start. 

\paragraph*{Non-Manifold Meshes}
Since Htex builds upon halfedge meshes, it is fundamentally limited to manifold meshes because 
halfedges can not describe non-manifold topologies. Hence, non-manifold meshes can not be used 
with Htex. We believe that a way to extend support to non-manifold topologies would be to rely 
on a mesh-matrix data-structure~\cite{DiCarlo2014,Zayer2017,Mahmoud2021}, but we have not looked 
into this. 
\section{Conclusion}
Htex is a novel parameterization-free texturing method that can be implemented 
on modern GPU. The key advantage of Htex over Ptex and mesh colors is that it 
supports non-quad topologies seamlessly thanks to the powerful properties of 
halfedges. While Htex is probably too costly to be used for an entire video game, 
we believe it could be used for assets that can benefit from crack-free 
displacement mapping with tessellation shaders. In future work, we would like to 
combine our method with halfedge-based subdivision algorithms~\cite{Dupuy2021} to render 
Ptex-based assets more faithfully within a unified, halfedge based, framework.

\section*{Acknowledgements}
The T-Rex asset from Figure~\ref{fig_teaser} was modeled by Olivier Roos 
for Studio Manette.
We thank Ignacio Llamas and Ignacio Casta\~no for sharing with us the 
asset from Figure~\ref{fig_dmap}.
Finally, we thank Brent Burley and Disney for giving access to the other 
T-Rex model from Figure~\ref{fig_ptex}.

\bibliographystyle{ACM-Reference-Format}
\bibliography{htex-bibliography}

\end{document}